\documentclass[10pt, conference, compsocconf]{IEEEtran}

\ifCLASSINFOpdf
  \usepackage[pdftex]{graphicx}
\else
\fi
\usepackage[cmex10]{amsmath}

\usepackage{bibspacing}
\usepackage{amssymb}
\usepackage{multirow}
\usepackage[noadjust]{cite}

\usepackage{tikz}

\newcommand{\ie}{i.e.}

\begin{document}
%
\title{Explore Recurrent Neural Network for PUE Attack Detection in Practical CRN Models}


\author{\IEEEauthorblockN{Qi Dong${^\dagger}$, Yu Chen${^\dagger}$, Xiaohua Li${^\dagger}$, Kai Zeng${^\ddagger}$}
\IEEEauthorblockA{${^\dagger}$Dept. of Electrical and Computing Engineering, Binghamton University, SUNY, Binghamton, NY 13902, USA\\
${^\ddagger}$Electrical and Computing Engineering Department, George Mason University, Fairfax, VA 22030, USA}
 E-mail: \{qdong3, ychen, xli\}@binghamton.edu, kzeng2@gmu.edu}


\maketitle

\begin{abstract}
The proliferation of the Internet of Things (IoTs) and pervasive use of many different types of mobile computing devices make wireless communication spectrum a precious resource. In order to accommodate the still fast increasing number of devices requesting wireless connection, more efficient, fine-grained spectrum allocation and sharing schemes are badly in need. Cognitive radio networks (CRNs) have been widely recognized as one promising solution, in which the secondary users (SUs) are allowed to share channels with licensed primary users (PUs) as long as bringing no interference to the normal operations of the PUs. However, malicious attackers or selfish SUs may mimic the behavior of PUs to occupy the channels illegally. It is nontrivial to accurately, timely detect such kind of primary user emulation (PUE) attacks. In this paper, an efficient PUE attacked detection method is introduced leveraging the recurrent neural network (RNN). After a fundamental algorithm using basic RNN, an advanced version taking advantage of long-short-term-memory (LSTM) is proposed, which is more efficient on processing time series with long term memory. The experimental study has provided deeper insights about the different performances the RNNs achieved and validated the effectiveness of the proposed detectors.
\end{abstract}

\vspace{8pt}


%
\IEEEpeerreviewmaketitle

\section{Introduction}

The rigid spectrum allocation scheme regulated by governmental agencies leads to great deficit on spectrum band resources. Under the current policies, wireless channels are assigned to fixed users for exclusive use. Such a static spectrum access technology results in a low efficient utility of wireless spectrum resources due to the fact that the frequency bands are scarcely used with geographical discrepancy. The emergence of new intelligent spectrum allocation/re-allocation schemes, especially cognitive radio networks (CRNs), are studied elaborately in the last decade, under the growing pressure from the ever-increasing wireless applications. Cognitive radio (CR), or known as secondary user (SU) in CRNs, is a technology that allows wireless devices (unlicensed users) access spectrum resources adaptively without introducing major interference to licensed primary users (PUs).

Basically, a well-designed CRN aims to serve for two purposes \cite{Adelantado:2013aa}: to maximize the usage of spare spectrum resource as well as to protect the incumbent primary system from secondary network interference. Since it is required to SUs that they shall not interfere the normal operations of the PU functionalities, SUs should adapt their behavior in accordance to PU activities. In most cases, knowing PU activities is essentially critical for cognitive radios to share the spectrum resource with legitimate users. One of the effortless ways to acquire PU activity information is that PUs are able to notify SUs their spectrum usage status; or there exist a third party as an inquiry center that knows what PUs will do in the near future. An alternative solution, which is the widely accepted one, is to develop robust and efficient spectrum sensing technique to acquire knowledge on PU activities.

Spectrum sensing allows CRs acquire real-time spectrum occupation information such that interweaving communications shared by PUs and SUs become feasible. Generally, spectrum sensing requires CRN participators to scan the desired spectrum band in a proper manner. In many CRN deployment models, especially interweave model, spectrum sensing is one of the most important procedure on exploring white spectrum space for use while imposing little interference to PUs \cite{Akhtar:2016aa}. Because of the limited computational capacity and energy resources of SUs, in many recent spectrum sensing and spectrum measurement studies, researchers had conducted intensive discussions on many practical implementations, such as sensing complexity, power consumption, sensing period, and so on \cite{Ali:2017aa}.

In practical CRN implementations, one of the major challenges of spectrum sensing is to detect PU signals with high accuracy while maintaining a low false alarm rate in the open and versatile radio environment, with limited SU resources. Also, the spectrum sharing efficiency greatly depends on a secure CR operating environment. In interweave spectrum sharing models, due to the opportunistic spectrum access (OSA) nature, CR systems encounter several CR-specified security problems. In contaminated radio environments, the false detection rate of spectrum sensing may become extraordinarily high, especially when a primary user emulation (PUE) attack happens. A PUE attack is that the malicious entities mimic PU signals in order to either occupy spectrum resource selfishly or conduct Denial of Service (DoS) attacks. PUE attacks can be easily implemented in CRNs. It introduces great overhead on cognitive radio communication and causes chaos in dynamic spectrum sensing. However, defense against the PUE attacks is nontrivial as traditional authentication and authorization (AA) methods are no longer applicable to CR systems, because no obligation should be imposed on PUs. More adaptive and practical PUE attack detection techniques are highly desired.

In general, PUE attack detection techniques are divided into two broad categories: fingerprint/radiometric based detection and activity pattern based detection. For the first category, the detection methods usually probe either a). the transmitter's intrinsic features such as pulse shape \cite{Kawalec:2004aa}, transient \cite{Rasmussen:2007aa}, b). the propagation channel features \cite{Huang:2010aa}, or c). geographical informations \cite{Chen:2009aa}. These fingerprint/radiometric based detection approaches all have restricted deployment due to particular limitations. For example, detection transmitter features are usually computing intensive, and some features are prone to be overwhelmed over long distance propagation. In addition, propagation channels are usually not static, and sometimes suffer from great fluctuation over short distances \cite{Chen:2008aa}. Also, geographical feature based detections usually require some critical prior informations, such as sensing nodes' location, PU propagation power, and stable propagation channel \cite{Chen:2009aa}. Furthermore, they are likely influenced by mobile PUs \cite{Yu:2015aa}.

On the other hand, activity pattern based methods have been discussed in several literatures \cite{Shan-Shan:2013aa, Xin:2014aa, Sharifi:2016aa, Dong:2017aa}. The fundamental idea is based on the assumption that PUs' activities can be generalized by some proper activity models or a combinations of a set of proper hyperplanes. When malicious nodes conduct PUE attacks, the sensed signal activities will show noticeable discrepancy from normal PU behaviors in compromised spectrum channels. Since attackers will inevitably generate excessive signal activities, activity pattern based detection methods are promising for development, due to the low computation burden and flexible deployment.

In most of the activity pattern based PUE attack detection studies, an un-intermittent spectrum scanning manner is assumed for each sensing SU, although intermittent spectrum sensing schemes are preferable in recently spectrum sensing studies, in seek of energy efficiency and processing efficiency \cite{Choi:2014aa, Xing:2014aa, Ali:2017aa}. In intermittent spectrum sensing, the sensing nodes scan the desired channel with a proper observation period to maintain a reasonable Resolution Bandwidth (RBW), and then sit back for a sensing interval to preserve energy and computation resource. Such kind of spectrum sensing schemes only allow CRN to obtain partial channel information, which brings great difficulties on detection of activity patterns.

When explore signal activity patterns, it is reasonable to assume that there are some ``features'' inherited into signal sequence. On the other hand, with an intermittent spectrum sensing, the channel status is not independent to previous status. Thus, the signal activity patterns can be regarded as the possible sequence which relates to some ``features'' of the channel and the previous internal states (memory). In this work, we proposed a recurrent neural network (RNN) based PUE attack detection method leveraging the energy-and-computation-efficient intermittent spectrum sensing technology. Further, a more advanced RNN techniques, called long short-term memory (LSTM), is studied for the detection.

The rest of this paper is organized as follows. Section \ref{sec:background} provides the background knowledge that motivated this work, it also discusses some related work on state-of-art spectrum sensing techniques and PUE attack detection approaches. Section \ref{sec:model} presents a practical PU activity model and briefly introduces the PUE attack detection process. The proposed RNN-based and LSTM-based PUE attack detection methods are introduced in Section \ref{sec:RNN}. Section \ref{sec:experiments} reports our experimental evaluation on three different PUE attack detectors in two PU activity models. Finally, Section \ref{sec:conclusions} concludes this paper along with a brief discussion about our on-going efforts.

\section{Related Work}
\label{sec:background}

\begin{figure}
	\centering
		\includegraphics[width=0.5\textwidth]{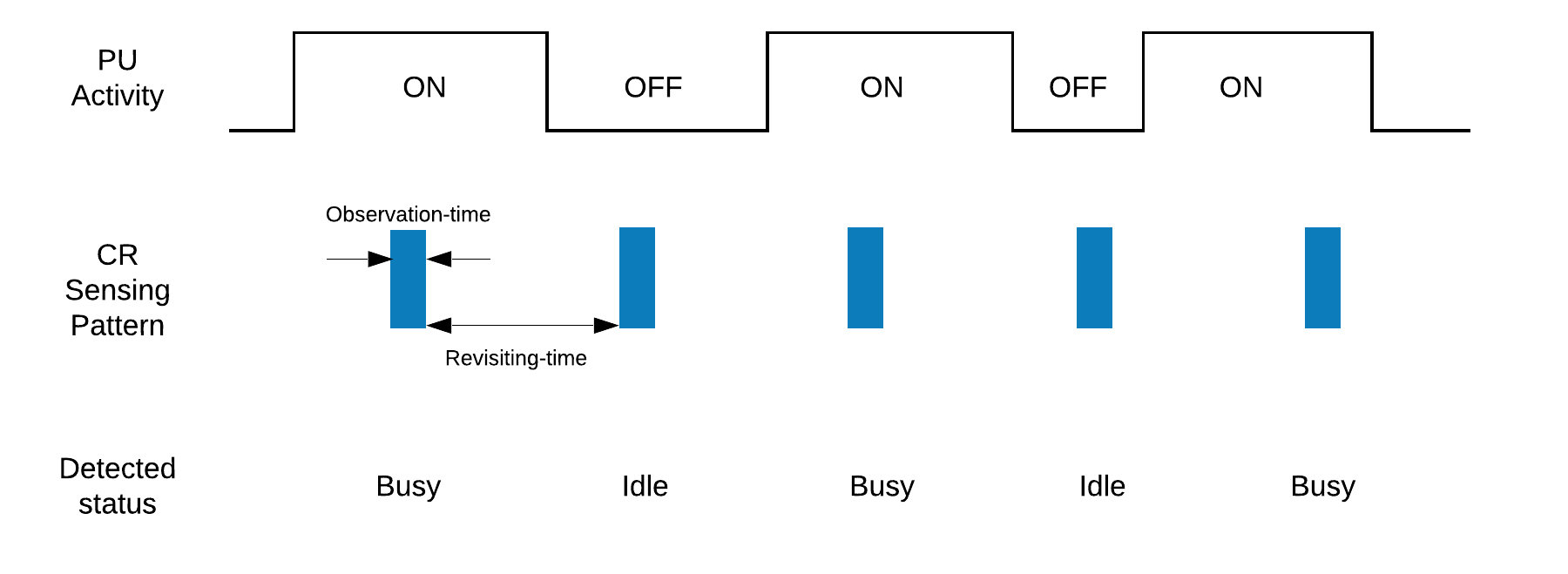}
	\caption{Practical CR spectrum sensing pattern.}
	\label{fig:ss}
    \vspace{-10pt}
\end{figure}

It is proven by many spectrum measurement campaigns around the world that the spectrum occupancy of PU channels shows noticeable statistic features, which can be described by some statistical models \cite{Chen:2016aa}. Figure \ref{fig:ss} shows a rational and practical way for spectrum sensing, which allows CRs measure a channel's state during a short observation time $T_{ob}$ and wait for a revisiting time $T_{re}$ until next spectrum sensing action. Based on such an intermittent spectrum sensing manner, there are two ways to improve the efficiency of CRNs. The CRs can either save energy during $T_{re}$ period of sleep, or they are able to perform wideband spectrum sensing by sensing other bands during $T_{re}$ period of scanning. However, because CRs only collect intermittent channel information, many signal patterns become untraceable. It is more advantageous for malicious parties to apply various attacks, especially the PUE attacks.

In CR systems, since no obligation is imposed on PUs according to Federal Communications Commission (FCC)  \cite{fcc2003}, it is necessary to distinguish attacker signals from PU signals while SUs actively sense the channel. Otherwise, PUE attacks will cause severe problems on the efficiency of spectrum utility \cite{Dong:2017bb}. As discussed before, in PUE attack detection, the fingerprint/radiometric based detection methods have restricted deployment, because they are either resource intensive, or only applicable under certain conditions. In comparison, activity pattern based detection methods attract more attention in broader application scenarios.

Some literatures have explored different ways using activity pattern for PUE attack detection. One of the recent work explored a cooperative spectrum sensing method on detection of PUE attack \cite{Sharifi:2016aa}. The author considered the most common spectrum sensing approach, energy detection, and find the optimal detection threshold of the aggregated sensing samples from all CRs by using numerical method. The proposed approach is easy to be implemented for its low cost, and it is also compatible to practical spectrum sensing process. However, the proposal is weak in several aspects: a) to find the optimal detection threshold, the prior probabilities of attack are assumed known; b) the sensed samples are assumed to follow the Gaussian distribution, which is not proven; and c) the attackers in the model are assumed too passive to adjust their behavior, such as attack strength, attack probability and attack timing. As a matter of fact, a smart attacker will easily deceive such detector.

In 2014, a signal activity pattern (SAP) based PUE attack detection method was proposed \cite{Xin:2014aa}. The main idea is that the PU signal activities can be well reconstructed by some sparse combination of a set of feature vector bases, while attack activities can not be sparsely reconstructed by such bases. Besides a lengthy training process, the proposal may suffer from two practical issues. Firstly, it requires a consecutive long enough observation duration to preserve ON/OFF features of the signal activities (30 ON/OFF periods in the literature). Secondly, it is expensive to find an optimal solution of the signal reconstruction problem by solving a convex optimization problem for each detection, which consumes a great amount of resources. Although the proposed SAP based PUE attack detection is theoretically sound, it is not advantageous in practical implementation.

Detection of PUE attacks greatly depends on the spectrum sensing manner. As discussed in many literatures, an efficient spectrum sensing method usually leverages only a short observation time on collecting signal samples \cite{Kim:2008ab, Choi:2014aa, Xing:2014aa, Cichon:2016aa}. The basis to find optimal spectrum sensing interval in CRN is that PU activities usually follow some activity models \cite{Saleem:2014aa, Chen:2016aa, Hoyhtya:2016aa}. In most of such activity models, each user request arrival is assumed independent, thus the state holding time or sojourn time is modeled as exponentially distributed \cite{Xing:2014aa, Rehmani:2013aa}. However, in practice, it was found that more complex distribution models are needed to better describe the state holding times, such as means of generalized Pareto or Hyper-Erlang distributions \cite{Benitez:2011aa, Fang:2001aa}.

To model a PU activity in spectrum sensing, various Markov models are preferably in many studies. A comprehensive study of the PU activity modeling by using first order Markov models can be found in \cite{Saleem:2014aa}. Although the first order Markov models may perform well for simple PU activity models, say exponentially distributed model, they are not adequate to model complicated, real-world PU activity models. Other studies, inspired by the great capability of artificial neural networks, made some trials on modeling PU activities using different neural networks (NN). For example, a plain feed-forward neural network (FNN) was used for spectrum prediction in \cite{Tumuluru:2010aa}. A simple FNN shows some merits on predicting spectrum status, but it omits one important fact. Similar to the first order Markov models, the next channel states are not only related to the current states, but also related to a possibly longer history of channel states. Recently, a trial by using a recurrent neural network (RNN) to predict spectrum slots is proposed \cite{Tang:2017aa}. A RNN is a neural network structure that makes use of sequential information by using its ``memory'' of previous states. It is naturally efficient on modeling the patterns of temporal dependency of time series \cite{Zheng:2017aa}.

\begin{figure}
	\centering
		\includegraphics[width=0.5\textwidth]{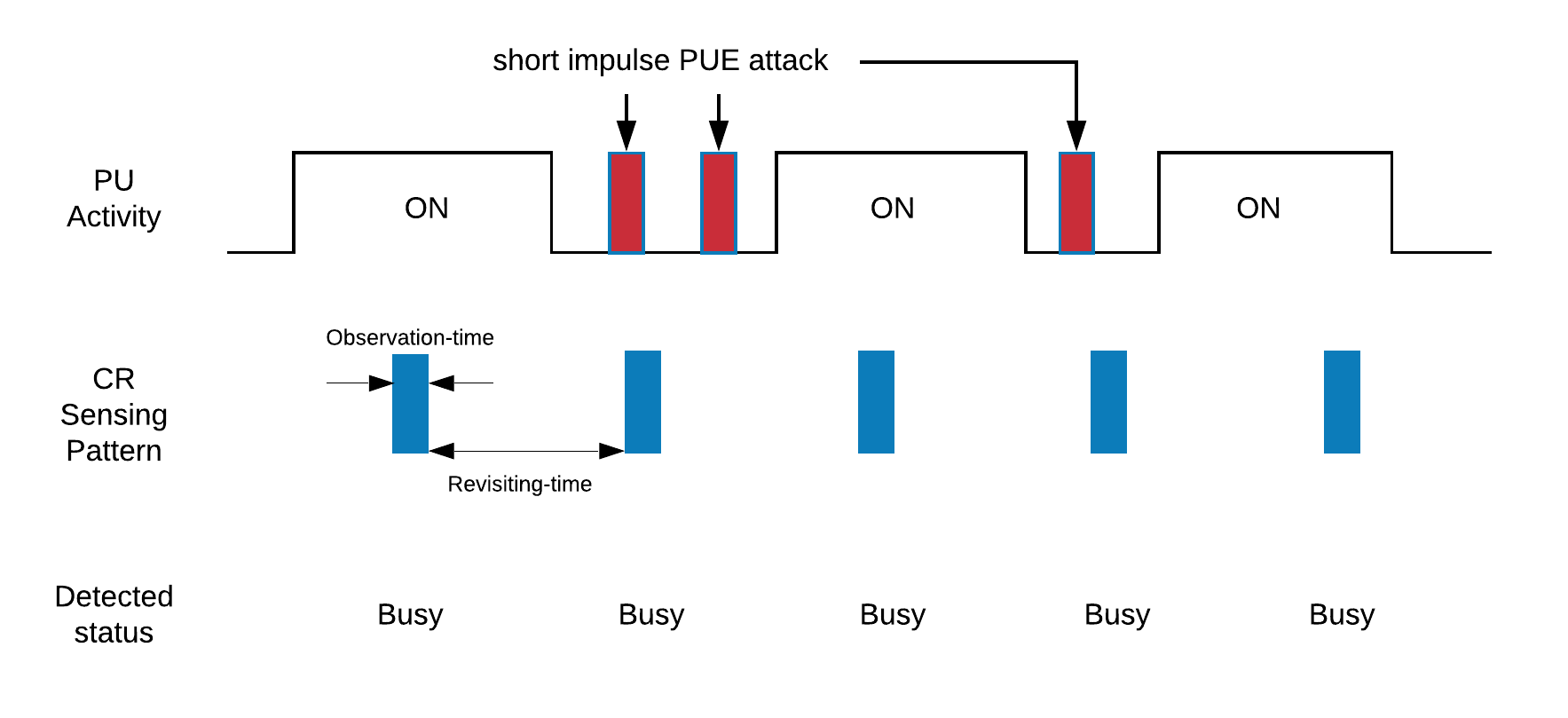}
	\caption{Short impulse PUE attack in CRN which uses intermittent spectrum sensing.}
	\label{fig:puea}
\end{figure}

While using the intermittent spectrum sensing approach, some critical PU activities information will get lost, such as channel holding time by PU. As a result, a smart attacker is able to conduct short impulse denial-of-service (DoS) attack to cause intensive PUE attacks as shown in Fig. \ref{fig:puea}. To overcome the partial information problem, in our proposal, we will explore the ``memory'' feature of the RNN to model the PU signal patterns of a spectrum channel, by which to detect the PUE attack in such channel.

\section{System model}
\label{sec:model}
In CRNs, the typical OSA is usually conducted by the following four steps: 
\begin{enumerate}
  \item apply spectrum sensing and analysis on wide spectrum band, and determine the proper bands for secondary use; 
  \item access a proper spectrum band for propagation while it is in idle state;
  \item apply intermittent spectrum sensing for consistent spectrum band monitoring, to avoid causing severe interference to PUs; and 
  \item switch to another proper spectrum band if current band is occupied.
\end{enumerate}

\begin{figure}
	\centering
		\includegraphics[width=0.5\textwidth]{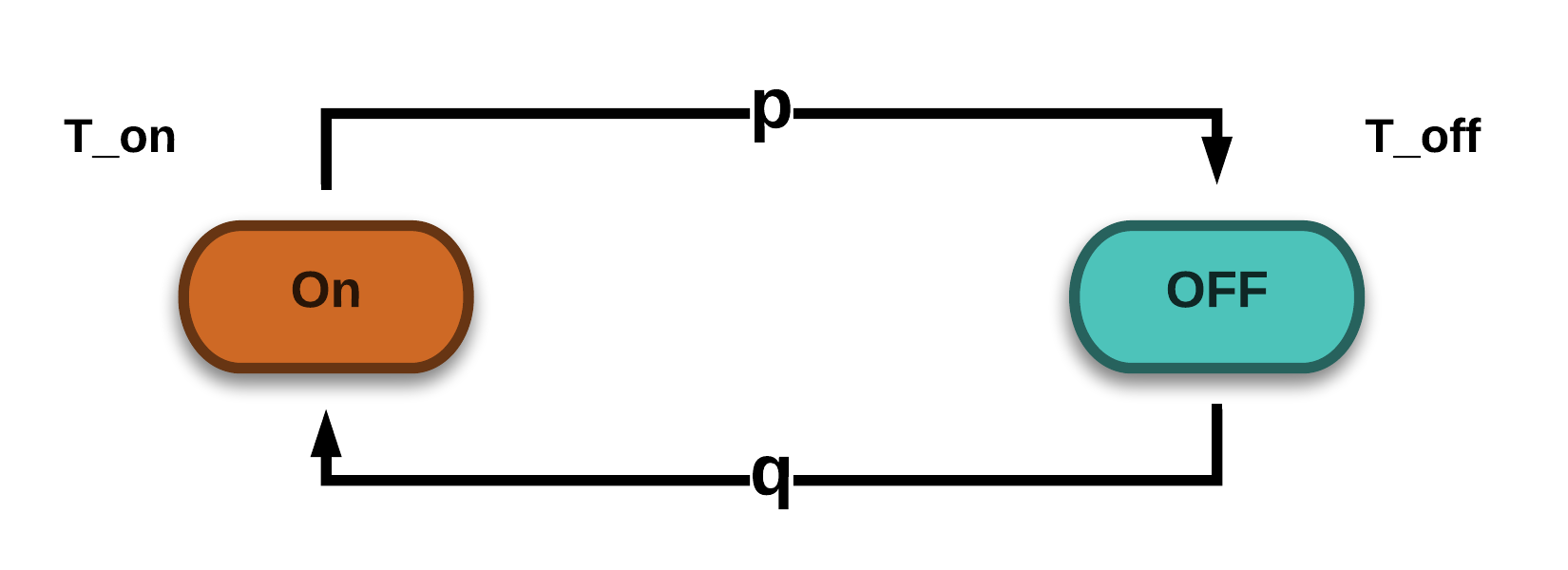}
	\caption{Two-state ON/OFF PU activity model.}
	\label{fig:onoff}
\end{figure}

As illustrated in Fig. \ref{fig:ss}, the PU activity in a channel can be described with a two-state ON/OFF model. In Fig. \ref{fig:onoff}, $p$ denotes the transition probability from ON state to OFF state, and $q$ denotes the transition probability from OFF state to ON state. $T_{on}$ stands for the sojourn time in ON state and $T_{off}$ denotes the OFF duration. The probability density functions (p.d.f) of $T_{on}$ and $T_{off}$ are denoted as $f_{T_{on}}(t)$ and $f_{T_{off}}(t)$, respectively.

For SUs using intermittent spectrum sensing, as discussed in previous section, CRs measure a channel's state during a short observation time $T_{ob}$ and wait for a revisiting time $T_{re}$ until next spectrum sensing action. During $T_{re}$, the SU can either put itself to sleep or sense other spectrum bands. One major problem with intermittent spectrum sensing is that the observation only reflects partial spectrum information. It senses ON or OFF at certain period, but missing some other critical information of the PU signals, such as the starting and ending moment of the ON and OFF period, and the exact length of ON and OFF period. Such an incomplete spectrum sensing measure makes it harder to detect malicious activities, especially the PUE attacks.

Fortunately, the PU activities are not totally random. It preserve some patterns and features. The involvement of the PUE attacks will inevitably introduce some ``other'' features. If we are able to learn the PU activity patterns from initial spectrum sensing and analysis stage, it is possible to distinguish exotic PUE activities from normal PU activities. We assume the PU ON/OFF activities follow Hyper-Erlang distributions, which have been proven as an effective PU activity model \cite{Fang:2001aa}. It is worthy to note that in practical spectrum prediction and abnormal detection, the PU activity model is unknown, and may follow other distributions. We will not make any strong assumptions about PU activity model's distribution function in our detection process. It means that we will not use the assumed Hyper-Erlang distribution model in our detection models. Instead, the Hyper-Erlang distribution model is used to simulate actual PU activities, where the $f_{T_{on}}(t)$ and $f_{T_{off}}(t)$ are defined as :

\begin{equation}
\label{eq:hon}
f_{T_{on}}(t) = \sum_{i=1}^n p_{on,i} \cdot \frac{t^{(k_{on,i}-1)}exp(-t/\theta_{on,i})}{\theta^{k_{on,i}}_{on,i}(k_{on,i}-1)!}
\end{equation}

\begin{equation}
\label{eq:hoff}
f_{T_{off}}(t) = \sum_{i=1}^n p_{off,i} \cdot  \frac{t^{(k_{off,i}-1)}exp(-t/\theta_{off,i})}{\theta^{k_{off,i}}_{off,i}(k_{off,i}-1)!}
\end{equation}

\noindent where $p_{on,i}, p_{off,i} \in \mathbb{R}$ and $\sum_{i=1}^n p_{on,i} = 1$, $\sum_{i=1}^n p_{off,i}=1$, $k_{on,i}, k_{off,i} \in \mathbb{N}$ are shape parameters, $\theta_{on,i}, \theta_{off,i} \in \mathbb{R}$ are scale parameters.

The expectations of $T_{on}$ and $T_{off}$ can be computed by:

\begin{equation}
\mathbb{E}(T_{on}) = \sum_{i=1}^n p_{on,i} \cdot  k_{on,i} \cdot \theta_{on,i}
\end{equation}

\begin{equation}
\mathbb{E}(T_{off}) = \sum_{i=1}^n p_{off,i} \cdot  k_{off,i} \cdot \theta_{off,i}
\end{equation}

In our detection model, the PUE attack detection process is briefly described in four steps: 
\begin{enumerate}
  \item The SUs will follow the typical OSA intermittent spectrum sensing process by first determining an appropriate spectrum channel by collecting a sensed time sequence $\boldsymbol{X}=\{x^{(1)},x^{(2)},\cdots,x^{(n-1)},x^{(n)}\}$, where it is composed by $1's$ and $0's$; the $1$ represents busy state of the channel, and $0$ represents idle state of the channel; 
  \item Based on the sequence $\boldsymbol{X}$, SUs can learn the signal activity patterns of the channel to construct a PUE attack detector $\mathcal{D}$; 
  \item While sharing the channel with PUs, SUs shall still apply intermittent spectrum sensing with the same manner to generate testing time sequence $\boldsymbol{S}=\{s^{(1)},s^{(2)},\cdots,s^{(n-1)},s^{(n)},\cdots\}$, in order to avoid interference with PUs and to detect malicious signal activities; and
  \item Use the current state series $\boldsymbol{s}^\circ=\{s^{(t-l_I)},s^{(t-l_I+1)},\cdots,s^{(t-1)},s^{(t)}\}$ with length $l_I$ as the detector $\mathcal{D}$'s input, and compare the output series $\boldsymbol{y}$ with the next state series $\boldsymbol{s}^{\ast}=\{s^{(t^\prime-l_C)},s^{(t^\prime-l_C+1)},\cdots,s^{(t^\prime-1)},s^{(t^\prime)}\}$ with length $l_C$ to determine the abnormity.
\end{enumerate}

\section{RNN based Detection}
\label{sec:RNN}
For activity pattern based PUE attack detection, the essential rationale is to extract or model the patterns of the normal signal traffic. With partial information series generated by intermittent spectrum sensing, the general idea of the detection becomes to determine whether or not the received series falls into the normal series set. Since the attacker can produce various PUE attacks, it is hard to define the attacker behavior, \ie~abnormal series set. A dilemma in such a situation is that it is not practical to construct a normal series set based on discrete sensing series whereas lacking of abnormal signal patterns. Because the longer series we use to construct the normal series set, the better to include more comprehensive normal behavior series in the set, while exclude the possible abnormal behavior series. The problems is, however, it will cause the constructed normal series set grows exponentially as the length of series grows. Another problem is that even if we can construct a normal series set, it still omits the internal connection between series states. For example, some behaviors of the series might be considered normal in some situations, while they are actually abnormal in different context. 

Thus, traditional approaches such as the signal decomposition, \ie~principal component analysis (PCA), or traditional classification techniques, \ie~vector machines and logistic regression, are not adequate to resolve PUE attack detection problem in this case. A more appropriate approach is to identify a relatively small normal series set, given a long history of previous behavior of series. For different history of series, the normal set shall change in a proper way. Therefore, a appropriate PUE attack detector with intermittent spectrum sensing should be: 1) not computational and memory expensive; 2) effective to preserve the PU activity patterns without any prior model information; and 3) normal pattern behavior is adaptive to a long history of behavior by seeking temporal domain information.

\subsection{RNN-based PUE Attack Detector}
Recurrent neural network (RNN), a feed-forward neural network (FNN) with decoupling temporal relationships and information, is a promising technique for the PUE attack detection. Similar to FNNs, the RNN is benefited from the adaptive structure that is able to model some complex nonlinear functions. In addition, RNN not only learns the network structure to fit the series model, but also preserve it cell states, which is comparable to the ``memory'' of the series while detecting. The current ``memory'' will be used along with the input series to predict the next step states. 

\begin{figure}
	\centering
		\includegraphics[width=0.3\textwidth]{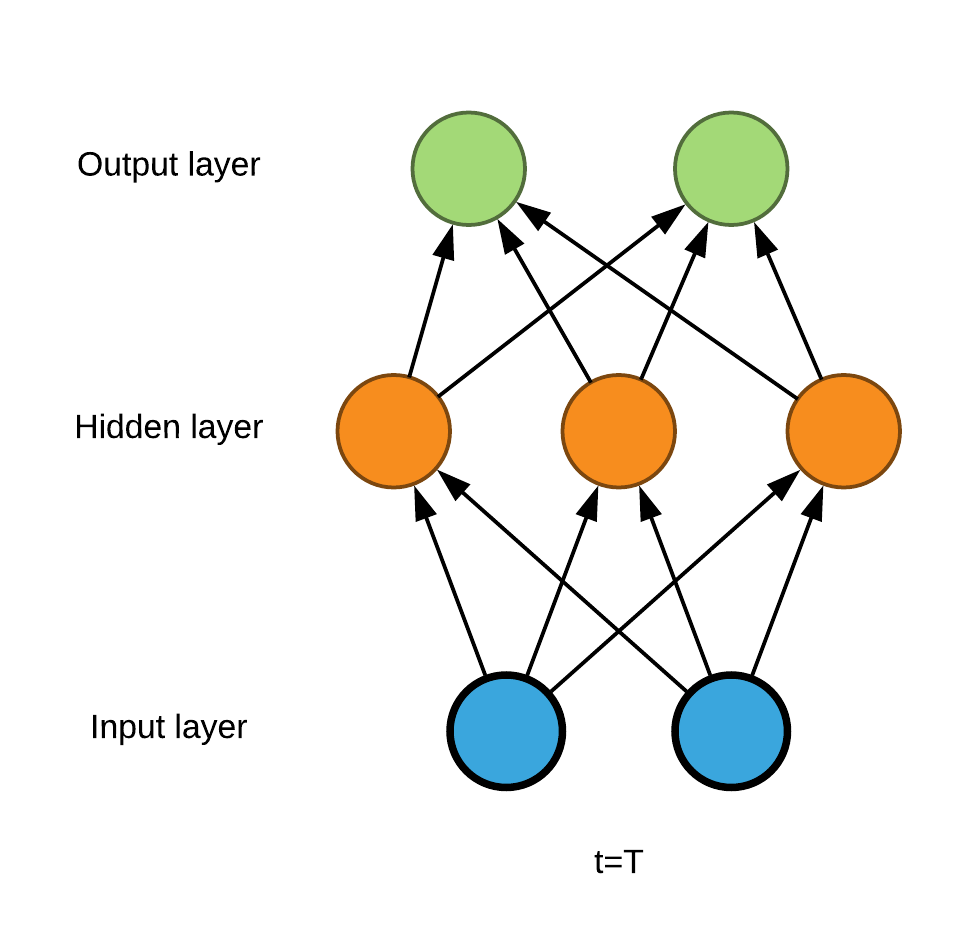}
	\caption{A simple feedforward neural network.}
	\label{fig:fnn}
\end{figure}

\begin{figure}
	\centering
		\includegraphics[width=0.5\textwidth]{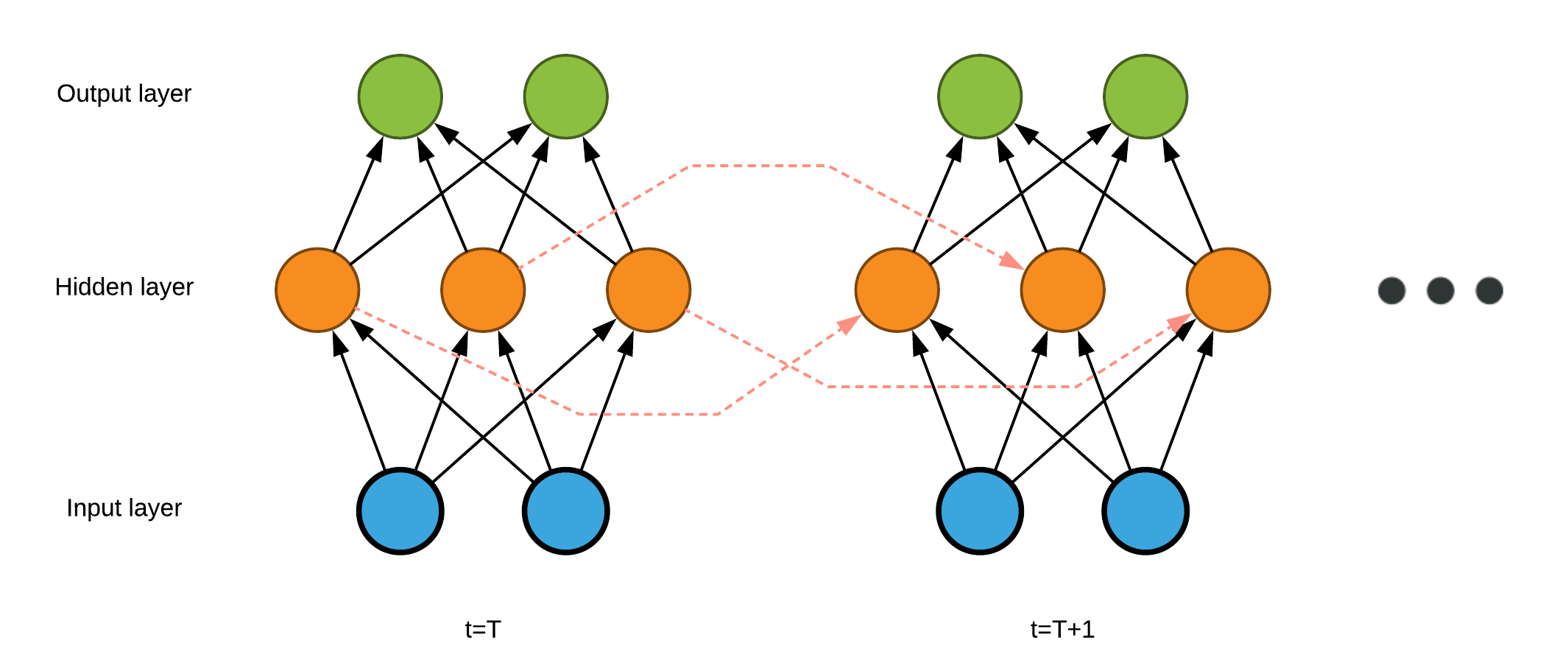}
	\caption{A simple recurrent neural network that unfolded across time steps. At each time step, the network uses both current input series and its cell states to generate the output.}
	\label{fig:rnn}
\end{figure}

Figures \ref{fig:fnn} and \ref{fig:rnn} show the different architectures of the basic FNN and RNN. In a FNN, the network only uses current time series to generate the output. The actual information used by the network only depends on the length of the input. It is problematic in series prediction, specifically in PUE attack detection. Because such a network structure generates uncertainty of which portion of the input information is more important and which portion is relatively trivial. Also it requires a large data set for training, since the input length is required to be sufficiently long. The RNN, on the other hand, not only uses current input series information, but also uses its memory of earlier time series in a recurrent way. Given a input series $\boldsymbol{S}=\{\boldsymbol{s}_1^\circ,\boldsymbol{s}_2^\circ,\cdots,\boldsymbol{s}_T^\circ\}$, the RNN computes the hidden state sequence $\boldsymbol{H} = \{\boldsymbol{h}_1,\boldsymbol{h}_2,\cdots,\boldsymbol{h}_T\}$, as well as the output sequence $\boldsymbol{Y} = \{\boldsymbol{y}_1,\boldsymbol{y}_2,\cdots,\boldsymbol{y}_T\}$ iteratively using the following set of equations, where $\boldsymbol{s}_i^\circ=\{s^{(t_i-l_i)},s^{(t_i-l_i+1)},\cdots,s^{(t_i-1)},s^{(t_i)}\}$ with length $l_I$, $\boldsymbol{h}_i=\{h_{(1)},h_{(2)},\cdots,h_{(m)}\}$, and $\boldsymbol{y}_i=\{y_{(1)},y_{(2)},\cdots,y_{(l_O)}\}$ with length $l_O$:

\begin{equation}
\label{eq:rnnc}
\boldsymbol{h}_i = f(\boldsymbol{W}_{hx}\boldsymbol{s}_i+\boldsymbol{W}_{hh}\boldsymbol{h}_{i-1}+\boldsymbol{b}_h)
\end{equation}

\begin{equation}
\label{eq:rnny}
\boldsymbol{y}_i = g(\boldsymbol{W}_{yh}\boldsymbol{h}_i+\boldsymbol{b}_y)
\end{equation}

In Eq. \ref{eq:rnnc} and Eq. \ref{eq:rnny}, $\boldsymbol{W}_{hx}$ is the matrix of conventional weights between the input and the hidden layer, $\boldsymbol{W}_{hh}$ is the matrix of recurrent weights between the hidden layer and itself at adjacent time steps, and $\boldsymbol{W}_{yh}$ is the hidden-output weight matrix. $\boldsymbol{b}_h$ and $\boldsymbol{b}_y$ are bias of the hidden layer and the output layer, respectively \cite{Lipton:2015aa}. $f$ is the activation function used in the hidden layer, and $g$ is the activation function used in the output layer.

In CRNs, during the initial spectrum sensing stage, the normal PU activity series can be used to train the RNN model at either SUs side or at a fusion center if a non-distributed CRN is adopted. While sharing the spectrum channel with PUs, the SUs can constantly inspect the signal activity behaviors, by applying the trained RNN as the PUE attack detector.

However, while training the RNN to learn PU activity patterns, we actually do not use any series with PUE attacks. The RNN can only learn the normal features. Thus, it is not able to classify the incoming activity series as either PU activity or attacker activity. We need to transform the trained RNN to an efficient RNN based detector $\mathcal{D}$. Since the RNN is modeling the features of the series composed by $1's$ and $0's$, the detector is designed to detect the difference between the predicted $\boldsymbol{y}_i=\{y_{(1)},y_{(2)},\cdots,y_{(l_O)}\}$ and the real received series $\boldsymbol{s}^{\ast}=\{s^{(t^\prime-l_C)},s^{(t^\prime-l_C+1)},\cdots,s^{(t^\prime-1)},s^{(t^\prime)}\}$ with length $l_C$, given some input series $\boldsymbol{s}^\circ=\{s^{(t-l_I)},s^{(t-l_I+1)},\cdots,s^{(t-1)},s^{(t)}\}$ with length $l_I$ and previous historical series. Actually, we do not expect the trained RNN to predict the exact output series. Instead, the series with length $l_C$ is mapped to a ``label'' domain with $2^{l_C}$ elements. For example, a series with length $2$ can be mapped to the ``label'' domain with $4$ elements, $\{[0,0], [0,1], [1,0,], [1,1]\}$. The RNN is trained to predict the likelihood of each different labels. Such that we are able to map the actually received series $\boldsymbol{s}^{\ast}$ to the element in the ``label'' domain. Then compute the mean square error (MSE) as the loss, by comparing predicted likelihoods of each labels and the real received series:

\begin{equation}
\label{eq:loss}
Loss = \frac{1}{2^{l_C}}\sum_{i\neq j} (y_{(i)})^2 + (1-y_{(j)})^2
\end{equation}

\noindent where $j$ denotes the label corresponding to the received series, $y_{(i)}$ denotes the output likelihood of label $i$.

\subsection{LSTM based detector architecture}
The basic RNN is not efficient in processing series with long temporal dependency, due to the well known gradient vanishing problem \cite{Bengio:1994aa}. The gradient vanishing problem makes RNN prone to forget long term information, which mean if some pattern happens long before current input series, RNN will not be able to use such pattern. In order to overcome this shortcoming, several other recurrent network structures were introduced. Long short-term memory (LSTM) is one of the important developments of RNN, in which the ordinary node in the hidden layer is replaced by a memory cell and several gates, shown in Fig. \ref{fig:lstm}. Each LSTM cell has two internal states: hidden state and cell state. In our case, given a input series $\boldsymbol{s}_i^\circ$, and previous hidden state $\boldsymbol{h}_i$, compute the forget gate $\boldsymbol{f}_i$ as:

\begin{figure}
	\centering
		\includegraphics[width=0.3\textwidth]{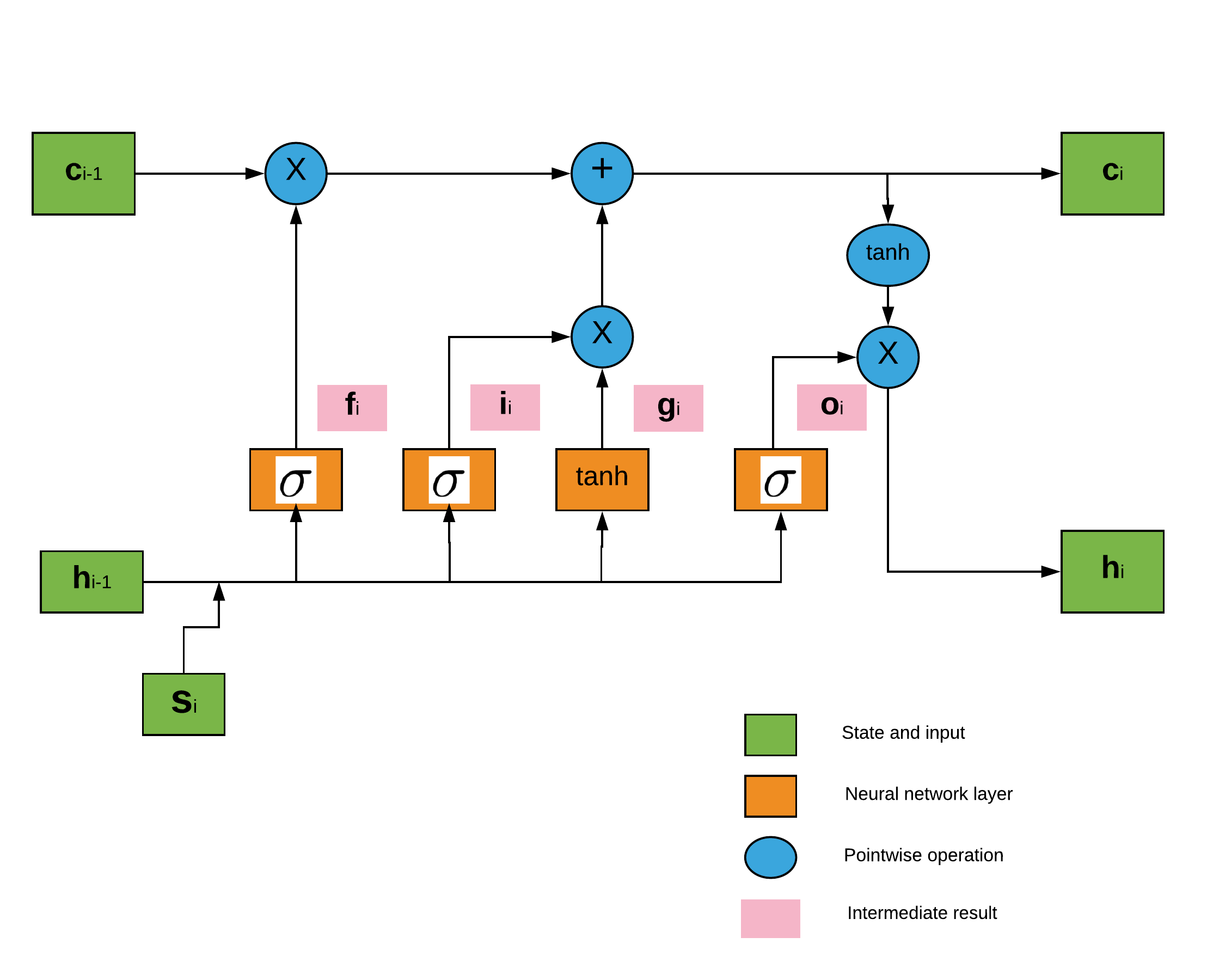}
	\caption{Structure of a Long Short-Term Memory cell.}
	\label{fig:lstm}
\end{figure}

\begin{equation}
\label{eq:forget}
\boldsymbol{f}_i = \sigma(\boldsymbol{W}_{fs}\boldsymbol{s}_i^\circ+\boldsymbol{W}_{fh}\boldsymbol{h}_{i-1}+\boldsymbol{b}_f)
\end{equation}

The output of Eq. \ref{eq:forget} is a matrix with elements between $0$ and $1$. It is used to decide what ``memory'' (cell state) is forgotten and what is kept, where $1$ means the memory is completely kept, and $0$ means the memory is totally forgotten. Further, the input gate $\boldsymbol{i}_i$ in Fig. \ref{fig:lstm} is computed as:

\begin{equation}
\label{eq:in}
\boldsymbol{i}_i = \sigma(\boldsymbol{W}_{is}\boldsymbol{s}_i^\circ+\boldsymbol{W}_{ih}\boldsymbol{h}_{i-1}+\boldsymbol{b}_i)
\end{equation}

The output of Eq. \ref{eq:in} is another matrix with elements between $0$ and $1$. It used to decide what input information should be used to update the ``memory'' (cell state), where $1$ stands for using the entire input information, and $0$ stands for not using the input information. The LSTM will process the input information $\boldsymbol{g}_i$ as:

\begin{equation}
\label{eq:inn}
\boldsymbol{g}_i = tanh(\boldsymbol{W}_{gs}\boldsymbol{s}_i^\circ+\boldsymbol{W}_{gh}\boldsymbol{h}_{i-1}+\boldsymbol{b}_g)
\end{equation}

With computed $\boldsymbol{h}_i$, $\boldsymbol{i}_i$ and $\boldsymbol{g}_i$, the new ``memory'' (cell state) is updated as:

\begin{equation}
\label{eq:cell}
\boldsymbol{c}_i = \boldsymbol{f}_i \odot \boldsymbol{c}_{i-1} \oplus \boldsymbol{i}_i \odot \boldsymbol{g}_i
\end{equation}

\noindent where $\odot$ means element-wise multiplication, and $\oplus$ means element-wise addition. Another gate called output gate $\boldsymbol{o}_i$ is used to decide what information should be used in new cell state as the output of this LSTM cell. The output gate $\boldsymbol{o}_i$ is computed as:

\begin{equation}
\label{eq:out}
\boldsymbol{o}_i = \sigma(\boldsymbol{W}_{os}\boldsymbol{s}_i^\circ+\boldsymbol{W}_{oh}\boldsymbol{h}_{i-1}+\boldsymbol{b}_o)
\end{equation}

The new hidden state as well as the output of this LSTM is computed as:

\begin{equation}
\label{eq:hidden}
\boldsymbol{h}_i = tanh(\boldsymbol{c}_i) \odot \boldsymbol{o}_i
\end{equation}

The hidden state of this LSTM cell, or the hidden state of the last LSTM cell for multi-layer LSTM, is fed into another activation function for final classification:

\begin{equation}
\label{eq:class}
\boldsymbol{y}_i = h(\boldsymbol{W}_{yh}\boldsymbol{h}_i+\boldsymbol{b}_y)
\end{equation}

\noindent where $h$ is the activation function for classification. In Eq. \ref{eq:forget} to \ref{eq:class}, $\boldsymbol{W}_{fs}, \boldsymbol{W}_{fh}, \boldsymbol{W}_{is}, \boldsymbol{W}_{ih}, \boldsymbol{W}_{gs}, \boldsymbol{W}_{gh}, \boldsymbol{W}_{os}, \boldsymbol{W}_{oh}, \boldsymbol{W}_{yh}$ are weight matrices, and $\boldsymbol{b}_f, \boldsymbol{b}_i, \boldsymbol{b}_g, \boldsymbol{b}_o, \boldsymbol{b}_y$ are bias vectors.

By applying above techniques in LSTM cell, the LSTM based PUE attack detector is theoretically more adaptive to both long term and short term of PU activity patterns than RNN based PUE attack detector.

In addition, the single layer LSTM can be extended with deeper structures, such as multi-layer LSTM. Theoretically, a multi-layer LSTM is able to learn more comprehensive PU activity patterns.

\section{Experimental Study}
\label{sec:experiments}
In this section, we will compare the performances of the basic RNN based detector, the LSTM based detector and the multi-layer LSTM based detector on PUE attack detections.

\subsection{Evaluation using a simple PU activity model}

We first evaluate the detectors performance using a relatively simple Hyper-Erlang distribution to model the PU activity. The parameters defined in Eq. \ref{eq:hon} and Eq. \ref{eq:hoff} are shown in Table \ref{tab:simple}.

\begin{table}[t]
	\caption{Parameters for simple PU activity model}
	\centering
	\label{tab:simple}
	\begin{tabular}{p{1cm}|p{2.5cm}|p{2.5cm}}
		\hline		
			  & Hyper-Erlang (ON) & Hyper-Erlang (OFF)	\\
		\hline
		$p$ 	  & $[0.5,0.5]$ & $[0.5,0.5]$	\\
		\hline
		$k$	 & $[1,1]$ & 	$[2,4]$\\
		\hline
		$\theta$ & $[0.5,1.5]$ & $[2,1]$ \\
		\hline
	\end{tabular}
\end{table}

The expectation of $T_{on}$ and $T_{off}$ can be computed as $1$ second and $4$ seconds respectively. We tentatively used a intermittent spectrum sensing interval $T_{re} = 0.24$ second and observation time $T_{ob} = 0.01$ second for spectrum sensing. In the contaminated PU activity, the attacker will deploy short impulse PUE attack with probability of $0.3$ at any particular time slot $T_{ob}$.

Table \ref{tab:lossSimple} shows the loss of using different RNN on predicting both normal series and abnormal series. The loss function is defined in Eq. \ref{eq:loss}. The statistic shows that the average loss on the attack series is much higher than the average loss on normal series.

\begin{table}[t]
	\caption{Average loss for different detector in simple PU activity model}
	\centering
	\label{tab:lossSimple}
	\begin{tabular}{p{2.5cm}|p{2cm}|p{2.5cm}}
		\hline		
		Method   & Average loss for normal series & Average loss for contaminated series 	\\
		\hline
		Basic RNN	 & $0.0235$ & 	$0.0538$\\
		\hline
		Single layer LSTM & $0.0491$ & $0.0901$ \\
		\hline
		Three layer LSTM & $0.0003$ & $0.0477$ \\
		\hline
	\end{tabular}
\end{table}

\begin{figure}
	\centering
		\includegraphics[width=0.4\textwidth]{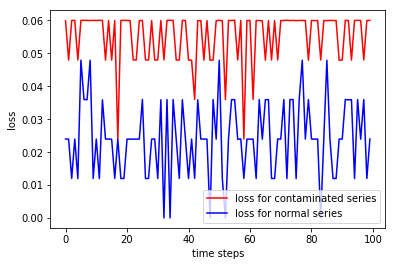}
	\caption{Sample of prediction loss using basic RNN in simple PU activity case.}
	\label{fig:rnn_simple}
\end{figure}

\begin{figure}
	\centering
		\includegraphics[width=0.4\textwidth]{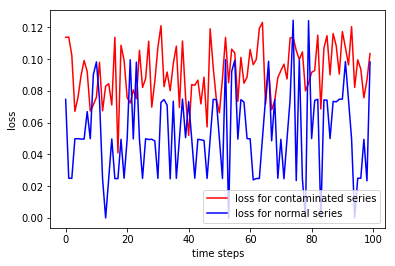}
	\caption{Sample of prediction loss using single layer LSTM in simple PU activity case.}
	\label{fig:lstm_simple}
\end{figure}

\begin{figure}
	\centering
		\includegraphics[width=0.4\textwidth]{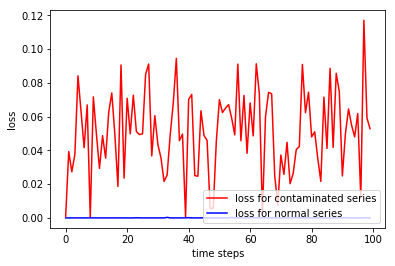}
	\caption{Sample of prediction loss using three layer LSTM in simple PU activity case.}
	\label{fig:mlstm_simple}
\end{figure}

Figures \ref{fig:rnn_simple}, \ref{fig:lstm_simple}, and \ref{fig:mlstm_simple} provide a closer look at how does the loss distributed, showing the loss distribution along a short period of time steps. All three figure show that the prediction loss on PUE attack contaminated series is almost always greater than the prediction loss on normal series. While using the contaminated series for prediction, on one hand, the output is highly unpredictable because the input series pattern may never be trained by the network; on the other hand, the correlation between the input series and real received series is obscure to the trained network. These two reasons cause the high loss of the contaminated series.

\begin{figure}
	\centering
		\includegraphics[width=0.4\textwidth]{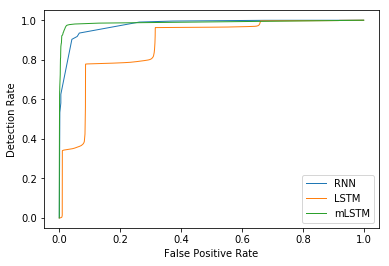}
	\caption{ROC curve of PUE attack detection in simple PU activity case.}
	\label{fig:roc_simple}
\end{figure}

Further, we use the output loss to plot the receiver operating characteristic (ROC) curve if use it for detection, shown in Fig. \ref{fig:roc_simple}. It shows that the three layer LSTM based PUE attack detector has the best detection performance, while the single layer LSTM based detector has the worst performance. It happens might because the our designed single layer LSTM is not able to learn the comlex feature with adequate capability, due to its shallow structure, while basic RNN only learns the short term features with better performance. The three layer LSTM detector achieves the optima performance because it is able to learn complex features as well as to preserve long term historical behaviors.

\subsection{Evaluation using a complex PU activity model}
In this subsection, we model the PU activity with a more complex form. The parameters are shown in Table. \ref{tab:complex}. In this case, the expectation of $T_{on}$ and $T_{off}$ can be computed as $4.296$ second and $8.23$ seconds respectively. We tentatively used a intermittent spectrum sensing interval $T_{re} = 0.99$ second and observation time $T_{ob} = 0.01$ second for spectrum sensing. In contaminated PU activity, we also set the attacker will deploy short impulse PUE attack with probability of $0.3$.

\begin{table*}[t]
	\caption{Parameters for complex PU activity model}
	\centering
	\label{tab:complex}
	\begin{tabular}{p{1cm}|p{7cm}|p{7cm}}
		\hline		
			  & Hyper-Erlang (ON) & Hyper-Erlang (OFF)	\\
		\hline
		$p$ 	  & $[0.2,0.05,0.1,0.1,0.2,0.05,0.1,0.03,0.07,0.1]$ & $[0.1,0.15,0.05,0.15,0.12,0.13,0.08,0.05,0.05,0.12]$	\\
		\hline
		$k$	 & $[2,1,2,2,1,3,10,4,3,6]$ & 	$[4,2,3,5,15,4,3,6,5,1]$\\
		\hline
		$\theta$ & $[0.5,1.2,0.3,0.6,2,0.8,1.2,1.8,2,2.5]$ & $[2.5,1.3,4,3,1,1.5,1,0.8,1.8,4]$ \\
		\hline
	\end{tabular}
\end{table*}

\begin{table}[t]
	\caption{Average loss for different detector in complex PU activity model}
	\centering
	\label{tab:lossComplex}
	\begin{tabular}{p{2cm}|p{2.5cm}|p{2.5cm}}
		\hline		
		Method   & Average loss for normal series & Average loss for contaminated series 	\\
		\hline
		Basic RNN	 & $0.0301$ & 	$0.0549$\\
		\hline
		Single layer LSTM & $0.0496$ & $0.0872$ \\
		\hline
		Three layer LSTM & $0.0001$ & $0.0478$ \\
		\hline
	\end{tabular}
\end{table}

\begin{figure}
	\centering
		\includegraphics[width=0.4\textwidth]{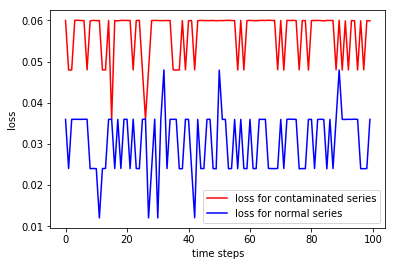}
	\caption{Sample of prediction loss using basic RNN in complex PU activity case.}
	\label{fig:rnn_complex}
\end{figure}

\begin{figure}
	\centering
		\includegraphics[width=0.4\textwidth]{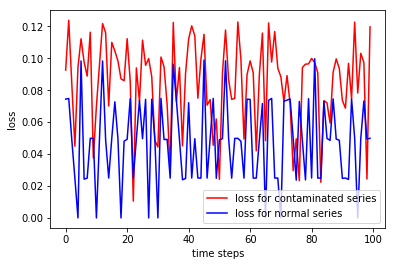}
	\caption{Sample of prediction loss using single layer LSTM in complex PU activity case.}
	\label{fig:lstm_complex}
\end{figure}

\begin{figure}
	\centering
		\includegraphics[width=0.4\textwidth]{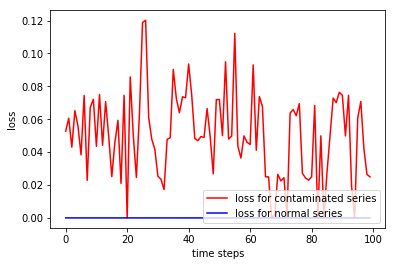}
	\caption{Sample of prediction loss using three layer LSTM in complex PU activity case.}
	\label{fig:mlstm_complex}
\end{figure}

\begin{figure}
	\centering
		\includegraphics[width=0.4\textwidth]{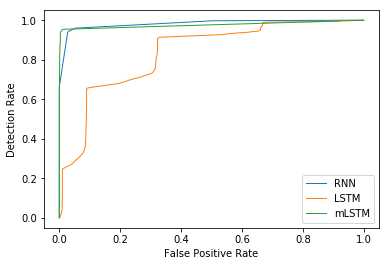}
	\caption{ROC curve of PUE attack detection in simple PU activity case.}
	\label{fig:roc_complex}
\end{figure}

Figures \ref{fig:rnn_complex}, \ref{fig:lstm_complex}, and \ref{fig:mlstm_complex} show the loss distribution along a period time steps. And Fig. \ref{fig:roc_complex} shows the ROC curve of different PUE attack detectors. Similar to the performance on simple PU activity model, the 3 layer LSTM based PUE attack detector has the best detection performance, while the single layer LSTM based detector has the worst performance.

If compare the performance of trained networks working in two different PU activity models, we will notice that both basic RNN based approach and single layer LSTM perform worse in complex PU activity model case (larger loss for normal series and smaller loss for contaminated series), while the three layer LSTM performs almost equally well in both cases. It proves that multi-layer LSTM can learn the complex PU activity features and utilize historical behaviors, thus performs better detection on PUE attack.

\section{Conclusion and Future Work}
\label{sec:conclusions}

This work aims at an effective PUE attack detection scheme. We first discussed the problem of PUE attack detection following a practical intermittent spectrum sensing approach. In order to achieve a better PUE attack detection rate on top of the partial information of the sensing series, a RNN based detection method is proposed, which is able to exploit temporal dependency of series for better series prediction and abnormal activity detection. Further, to overcome the gradient vanishing problem existing in the basic RNNs, we proposed a LSTM based detection method that can use longer terms of series pattern for better performance. Our experimental results shows that a multi-layer LSTM based PUE attack detector has achieved a superior performance.

In our experimental study, we tentatively chose some spectrum sensing intervals based on the expectation of $T_{on}$ and $T_{off}$. Although there are reported studies that discussed how to select an optimal interval to maximize the SUs benefits, it is advantageous to discuss the optimal sensing interval considering both SUs propagation as well as anomaly detection. Our experimental results showed that the single layer LSTM based detector performs not as good as the multi-layer LSTM based detector or even the basic RNN. It worthy a trail to develop more advanced LSTM based detectors. Newer LSTMs such as peephole LSTM, Gated Recurrent Units (GRU), might gain more efficiency on learning and show better performance on detection. Furthermore, we hope our work can be extended to use in real CRNs. Part of our on-going efforts also include identifying more real-world scenarios to deploy our proposed RNN based PUE attack detectors.

\section*{Acknowledgement}

Q. Dong, Y. Chen and X. Li are supported by the NSF via grant CNS-1443885. K. Zeng is partially supported by the NSF under grant No. CNS-1502584
and CNS-1464487.



%

\footnotesize

\bibliographystyle{IEEEtranS}
\bibliography{CRN_DQ}

\end{document}